\documentstyle[12pt,aaspp4]{article}
\def \yskip{\penalty-50\vskip3pt plus 3pt minus 2pt}
\def \pp{\par \yskip \noindent \hangindent .4in \hangafter 1}
\def \abc#1#2#3#4 {\pp#1, {\sl#2}, {\bf#3}, #4}
\def \blank {\lower 5pt\hbox to 0.75in{\hrulefill}}
\def\etal{{\it et al.}}

\newfont{\rten}{cmr10}

\def\arcsec{\hbox{$^{\prime\prime}$}}
\def\HST{{\it HST}}

\begin{document}

\normalsize
 
\title{Constraints on Cosmological Models from Hubble Space Telescope Observations of High-$z$ Supernovae}
\vspace*{0.3cm}

Peter M. Garnavich\footnote{Harvard-Smithsonian Center for Astrophysics, 60 Garden St., Cambridge, MA 02138},
Robert P. Kirshner$^1$,
Peter Challis$^1$,
John Tonry\footnote{Institute for Astronomy, University of Hawaii, Manoa},
Ron L. Gilliland\footnote{Space Telescope Science Institute, 3700 San Martin Drive, Baltimore, MD 21218},
R. Chris Smith\footnote{University of Michigan, Department of Astronomy, 834 Dennison, Ann Arbor, MI 48109-1090},
Alejandro Clocchiatti\footnote{Cerro Tololo Inter-American Observatory, Casilla 603, La Serena, Chile}$^,$\footnote{Current address: Pontificia Universidad Catolica de Chile, Departamento de Astronomia y Astrofisica, Casilla 104, Santiago 22, Chile},
Alan Diercks\footnote{Department of Astronomy, University of Washington, Seattle, WA 98195},
Alexei V. Filippenko\footnote{Department of Astronomy, University of California, Berkeley, CA 94720-3411},
Mario Hamuy\footnote{Steward Observatory, University of Arizona, Tucson, AZ 85721},
Craig J. Hogan$^7$,
B. Leibundgut\footnote{European Southern Observatory, Karl-Schwarzschild-Strasse 2, D85748 Garching, Germany},
M.M. Phillips$^5$,
David Reiss$^{7}$,
Adam G. Riess$^8$,
Brian P. Schmidt\footnote{Mount Stromlo and Siding Spring Observatory, Private Bag, Weston Creek P.O., ACT 261, Australia},
%Robert A. Schommer$^5$,
J. Spyromilio$^{10}$,
Christopher Stubbs$^{7}$,
Nicholas B. Suntzeff$^5$,
Lisa Wells{$^9$}

\begin{abstract}

We have coordinated {\it Hubble Space Telescope} photometry with ground-based discovery 
for three supernovae:
two SN~Ia near $z\sim 0.5$ (SN~1997ce, SN 1997cj) and a third event at $z=0.97$ (SN~1997ck).
The superb spatial resolution of \HST\ separates each supernova from its host galaxy and leads
to good precision in the light curves. The \HST\ data combined with ground-based photometry
provide good temporal coverage. We use these light curves and relations between 
luminosity, light curve shape, and color calibrated from low-$z$ samples to derive 
relative luminosity distances
which are accurate to 10\%\ at $z\sim 0.5$ and 20\%\ at $z=1$. The
redshift-distance relation is used to place constraints on the global mean 
matter density, $\Omega_m$, and the normalized cosmological constant, $\Omega_\Lambda$. 
When the \HST\ sample is combined
with the distance to SN~1995K ($z=0.48$), analyzed by the same precepts, it suggests
that matter alone is insufficient to produce a flat Universe. Specifically, for
$\Omega_m+\Omega_\Lambda =1$, $\Omega_m$ is less than 1 with $>95$\%\ confidence, and 
our best estimate of $\Omega_m$ is $-0.1 \pm 0.5$ if $\Omega_\Lambda =0$.
Although the present result is based on a very small sample whose systematics 
remain to be explored, it demonstrates the power of \HST\ measurements for high redshift supernovae.
\end{abstract}


\section{Introduction}

The direct measurement of global curvature and deceleration of the Universe
has challenged the best efforts of 
observers for many decades (Humason, Mayall and Sandage 1956, Baum 1957, Sandage 1988).
Progress has been stymied by a lack of reliable standard candles and yardsticks, and the
difficulty of making precise measurements on faint objects at high redshift. Two recent
advances now offer the hope of solving this classical problem:
the empirical calibration of Type Ia supernovae (SNe~Ia) as precise distance indicators, and
new technology that allows the measurement of supernova (SN) properties at large distances.
This {\it Letter} reports the results from coordinated ground-based and
{\it Hubble Space Telescope} (\HST) observations of
distant SNe up to $ z\sim1$ in an effort to extend the luminosity-distance
relation to regions where the cosmological effects of
deceleration and curvature can be measured (N\o rgaard-Nielsen et al. 1989;
Perlmutter et al. 1997; Schmidt 1997).

Type~Ia SNe are proven to be excellent distance indicators.
They are not perfect standard candles, but their luminosities correlate with
light curve shape so differences in intrinsic brightness can be taken into account
by observing the SN light curve  (Phillips 1993;
Hamuy \etal\ 1996 [H96]; Riess, Press, \&\ Kirshner 1996 [RPK]).
Using the techniques of RPK and H96, relative distance 
estimates to individual
events are accurate to better than 10\%\ despite a range of more than a 
magnitude in
intrinsic brightness. Colors of SNe~Ia provide a way to 
correct for extinction by dust in both our Galaxy and the host.  
Cosmological models predict the shape of the relation between luminosity 
distance and redshift, thus relative distances constrain curvature and
deceleration independent of the Hubble constant. By 
using the same methods at high and low redshift, based on our extensive study of SN Ia,
we expect to minimize the systematic
errors that can undermine any enterprise of this type.
Even evolutionary effects, which have bedevilled all previous
attempts to measure $q_0$, can to some extent be calibrated by
studying contemporary samples in populations of different ages
and metallicity; Schmidt et al. (1997a, [SSP97]) demonstrate that any current 
population-dependent bias in luminosity is less than $m-M=0.06$ mag. 

Our team has undertaken a program to discover and study SNe~Ia at $z>0.3$ 
(Schmidt \etal\ 1995, 1997b; Kirshner \etal\ 1995; Suntzeff \etal\ 1996;
Garnavich \etal\ 1996, 1997a, 1997b; Leibundgut \etal\ 1996;
Riess \etal\ 1997; Tonry \etal\ 1997).
We have discovered more than 70 candidates which include nearly 30 confirmed SN~Ia.
Results from our first discovery (SN~1995K at $z=0.48$), a detailed description
of our techniques, and a discussion of possible systematic errors are given by SSP97. Here, we
present preliminary results from our ground-based search combined with \HST\ and ground-based
photometry. \HST's exquisite imaging reduces the contribution from host galaxy light and \HST's
indifference to lunar phase provides continuous coverage over the 
period needed to define the shape of the light curve.
In \S 2 we discuss the discovery and photometry of two SNe~Ia at 
$z\sim 0.5$ and an object
at $z\sim 1$ which is very probably a type~Ia event. In \S 3 we analyze the 
light curves
using the Multicolor Light Curve Shape (MLCS) method (RPK) and a template 
fitting technique
(Hamuy \etal\ 1995) to determine distances. In \S 4, we
produce a consistent Hubble diagram in the range $0.01 < z < 1.0$ and use
the results to constrain cosmological models.

\section{Observations}
\subsection{The Search}
The \HST-coordinated search for distant SNe was conducted at the Canada-France-Hawaii 
Telescope (CFHT)
on 1997 April 29 and 30 (UT) using the UH8K CCD mosaic. The camera has eight 4096x2048 pixel
arrays and a field of view of nearly 0.5 degrees
with a scale of 0.21 arcsec/pixel.
Four fields, selected and scheduled long in advance for \HST\ visibility, were imaged in
broad-band $I$ and $V$ filters, for a total
search area of about one square degree.
We compared the images with template frames of the same
regions taken April 4 and 9 (UT) to detect variable objects. The seeing
on the search frames was between 0.5$''$ and 0.6$''$ and approximately 0.7$''$ (FWHM)
on the template frames. A minimum of three 1200 second exposures
were taken of each field and combined with a median filter.  This
process eliminates cosmic ray events, CCD flaws and asteroids. The magnitude limit 
of each field varied with exposure times and seeing, but typical 
isolated stellar images could be detected to $I < 24.5$.
In software, the two sets of images were aligned, convolved to match
point-spread functions, scaled, and subtracted. The subtracted images were then
searched in software for residual point sources and inspected by eye. 

Twelve SN candidates were identified from the search.  Two of the objects had
blue $V-I$ colors unlike any high-redshift SN~Ia. Spectra of eight
of the remaining candidates were obtained with the Multiple-Mirror Telescope on May 
1 and 2, and with the Keck 10m telescope on May 4.
One of the candidates had a spectrum characteristic of an active galactic 
nucleus, two could not be classified, one was a SN~II at $z=0.28$, and four of the objects
were identified as SN~Ia (Tonry et al., 1997).  We selected
SNe 1997ca, 1997ce, 1997cj and 1997ck for \HST\ observations based on
the \HST\ scheduling requirement to have exactly one target in each of the four fields and 
the likelihood that the target was a SN~Ia before maximum from the
spectrum and the photometry in hand on May 5. 
Subsequent photometry of SN~1997ca indicated that it probably is not a SN~Ia. A reanalysis of the 
spectrum shows that it is consistent with a SN~II at $z\sim 0.4$.   

\subsection{\HST\ and Ground-based Imaging}

Following discovery, photometry of the SNe was obtained with a variety of ground-based telescopes
and eventually with \HST. Even for a well-coordinated program, the interval between discovery and the 
first \HST\ observation can be more than two weeks so earth-based data are important to
define the light curve before maximum light.

Our first WFPC2 images were
obtained on 1997 May 12, just a week after the Keck spectra.
We observed each SN in the WF3 chip on six visits spanning
approximately three weeks in the SN rest frame. Each epoch was allotted
one \HST\ orbit. For the $z\sim 0.5$ targets, the orbit was
divided into a 800 second exposure in the F675W filter and a 1100 second
exposure with the F814W filter. These filters approximate the standard $B$ and $V$
bandpasses at $z\sim 0.5$ and we have computed K-corrections
for both the \HST\ and ground-based observations according to SSP97.
At $z\sim 1$, the F850LP is well matched to the
rest frame $B$ band while the $V$ band is shifted beyond the limits of WFPC2
sensitivity. The exposures of SN~1997ck in the F850LP filter were set to fill the target 
visibility window with a minimum total exposure of 2200 seconds.
All observations were divided into two exposures, and we combined the
cosmic-ray split images using the default parameters in the STSDAS/HST\_CAL/WFPC/CRREJ
algorithm which is designed to avoid confusing stellar images with cosmic rays in undersampled data.
As a test, we performed aperture photometry on bright, unsaturated stars observed at each epoch and found
a scatter of less than 0.01 mag, consistent with the predicted statistical error. 
Plate~1 displays images of each SN made by adding all the observations.

We calibrated a sequence of stars near each supernovae using both \HST\ and ground-based
data. The magnitudes of stellar objects in the \HST\ images were estimated using
the prescription of Holtzman \etal\ (1995). The data numbers within 0.3$''$
radius aperture were summed and we subtracted the background level estimated from a
large region around the image. The small aperture was selected to minimize background noise, so an
aperture correction based on the PSF created from stars in the field was applied
to bring the measurement to the equivalent of a 0.5$''$ aperture and the result
converted to a magnitude in the \HST\ filter system.  For the F850LP filter, only a synthetic zeropoint (ZP)
was available but it is estimated by Whitmore (1995) to be good to $\sim 3$\% .

Observations of standard stars and three of
the \HST\ fields were obtained in the $R$ and $I$ bands under photometric conditions on
three nights with the Hawaii 88-inch telescope in 1997 May. The F675W and F814W magnitudes for
15 stars in common with the Hawaii data were converted to $R$ and $I$ magnitudes using ZPs
and second order color terms provided by Holtzman \etal\ (1995). There is a
significant color residual between the two calibrations, but for the seven stars
with colors similar to those expected for the SNe ($R-I < 1.2$),
the average difference between the ground-based (GRD) and \HST\ magnitude estimates
is only $R_{GRD} - R_{HST} = -0.02\pm 0.05$ mag and $I_{GRD} - I_{HST} = 0.00\pm 0.04$ mag, 
verifying the Holtzmann ZPs.

From the ground, light from the host galaxy is a major source of uncertainty in measuring SN light curves. 
\HST\ allows the SN and the galaxy light to be more
easily separated, but the host background must still
be removed. For SNe~1997ce and 1997cj, an empirical point-spread-function (PSF) derived from
nearby stars in the field was scaled to the peak brightness of the SN and subtracted from each
observation. The four pixels at the SN position were smoothed, then all the epochs
combined to produce one high signal-to-noise ratio image. This template was
subtracted from each \HST\ epoch leaving only the SN and other stars in the image to be
measured with aperture photometry. The PSF subtraction is not perfect, and we have estimated,
through simulations, the uncertainty in the background (always $<2$\%), and included this
error with the estimated statistical error of each measurement.
The host galaxy for SN~1997ck was insignificant in the F850LP filter, so no correction was necessary.

Reduction of the ground-based data followed the procedures outlined by
SSP97.  The \HST\ image templates, described above, were used to
subtract the host galaxy background, and the relative brightness between the SN and
other stars in the field were estimated for each frame using
PSF fitting routines in DoPHOT (Schechter, Mateo, \& Saha 1993). Artificial stars
were added and measured to estimate the uncertainty of each measurement, 
with these errors added in quadrature to those resulting from the imperfect template.

In both cases, the ZP (and color term for the ground-based data) were determined
using the F675W and F814W magnitudes of the field stars on the \HST\ frame.
K-corrections were applied to both the \HST\ and ground-based data
to bring the F675W and F814W magnitude estimates to rest-frame $B$ and $V$ respectively. 

\begin{figure}[ht]
\plotfiddle{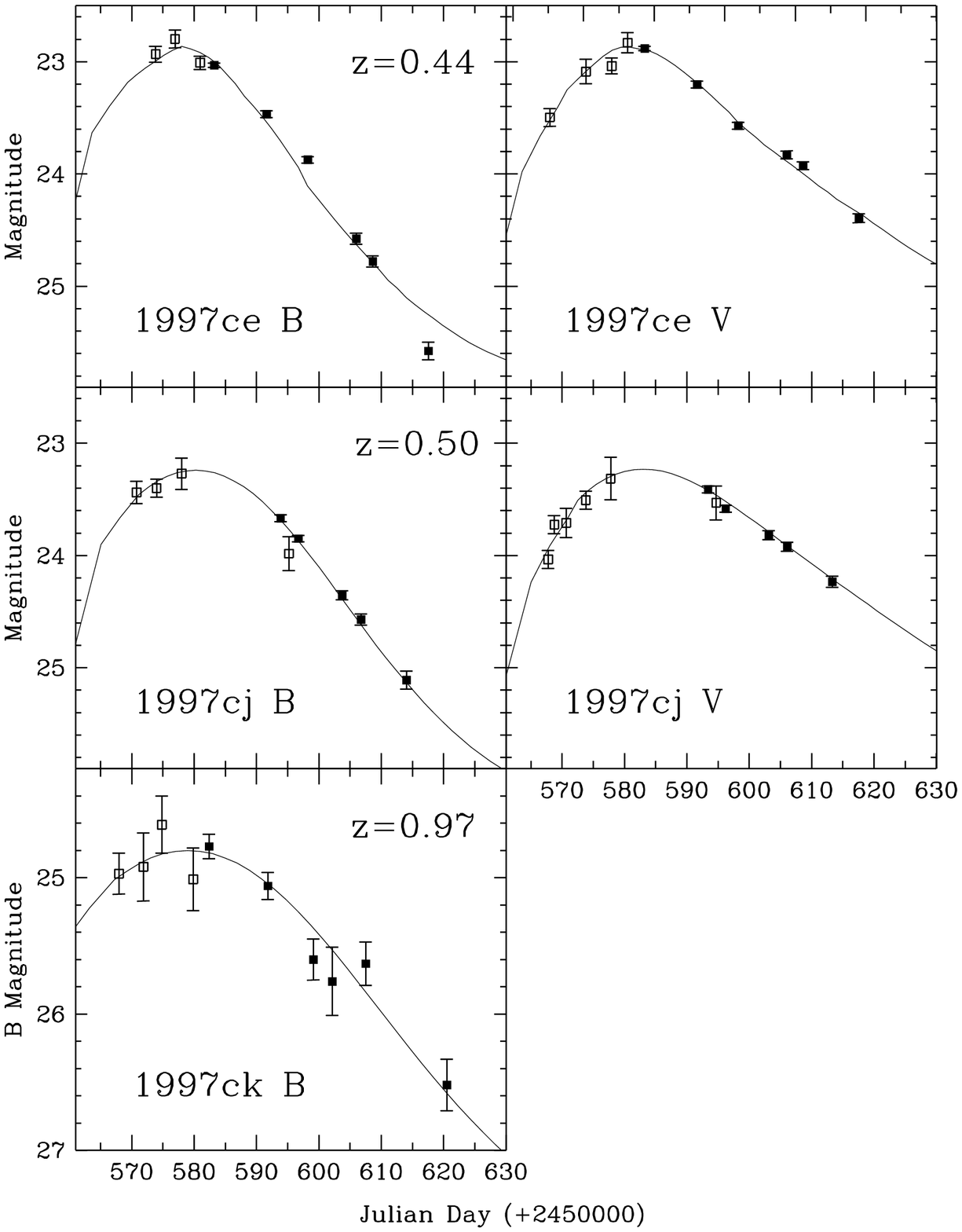}{14cm}{0.0}{80.0}{80.0}{-250.0}{-80.0}
\vspace*{1.0cm}
\caption{Rest frame $B$ and $V$ light curves of SNe~1997ce (top) and 1997cj 
             (middle) amd a $B$ light curve for SN~1997ck (bottom).
             The filled points represent \HST\ observations while the open points 
             are from ground-based telescopes.
             The solid lines show the best fit light curve from the MLCS method.}
\end{figure}

In addition to the statistical errors (shot-noise) in the SN magnitudes, there are
sources of possible systematic error which contribute to the SN distance estimates
and we list them in Table~1. The Holtzman calibration is expected to have
a zeropoint uncertainty of 3\%, but recent
comparisons of Holtzmann magnitudes with ground-based data suggests that ZP
offsets and uncertainties of 0.05~mag are possible, and errors of this magnitude
cannot be excluded with our current ground-based data. 
Inefficient charge transfer (CT) in the WF3 causes the apparent
brightness of objects to vary with pixel position. A recent characterization of the CT problem
for WFPC2 by Whitmore \&\ Heyer (1997) shows that the
loss of charge along columns can be as large as 7\%\ and depends on object brightness
and background level as well as pixel position. We applied their corrections to all the magnitude
estimates after interpolating to a 0.3$''$ aperture and this lowered the estimated SN magnitudes by
between 3\%\ and 5\%, with the corrections being good to about 2\%.
Tests by Hill \etal\ (1997) found differences in the WFPC2 ZP between long and short exposures
amounting to 5\%. However, these offsets are symptomatic of the CT problem and are corrected
in our data with the new algorithm.

\section{Results}

The \HST\ image shows that SN~1997ce occurred 0.4\arcsec\ south of
the brighter of a pair of elongated galaxies. 
Keck spectra of SN~1997ce displayed a blue continuum with
broad absorption bands which matched those of a SN~Ia at $z=0.44$.
The light curve showed that SN~1997ce was discovered about 8 days before
maximum light in the observer's frame (roughly 6 days in the rest frame).
The MLCS method was used to analyze the rest-frame light curves (Figure~1) and
found that SN~1997ce declined slightly faster than a normal SN~Ia.
Template fitting agreed, but found a smaller correction. The MLCS fit found
no reddening of this SN, consistent with the color at maximum of $B-V\approx0$.

The host galaxy of SN~1997cj is a spiral with
the SN offset by 0.7\arcsec\ to the west.
The MMT spectrum of SN~1997cj showed features consistent with
a SN~Ia at a redshift of 0.5 and a narrow emission line of [OII] $\lambda 3727$
provided a precise redshift of $z=0.50$. Discovery was about 12 observer days 
before maximum light.
Although the first planned \HST\ visit could not be made
due to a lack of guide stars, the light curves (Figure~1) are still well 
defined by combined ground-based and \HST\ data and
show SN~1997cj to have a normal decline rate. MLCS fits to the $B$ and $V$ light curves
argue that the object is not reddened.

The Keck spectrum of SN~1997ck was too
weak to show the broad features of the SN, but strong emission 
from the host galaxy corresponding to [OII] $\lambda 3727$ indicated a redshift of 
$z=0.97$. The $V-I$ color was consistent with a SN~Ia before maximum. The
host is not easily seen in the F850LP \HST\ images, but in the $R$ and $I$ 
ground-based frames it appears as a very elongated, low surface-brightness patch extending $\sim 
2$\arcsec\ to
the northwest. The blue color of the host may indicate a population of young 
stars and there is a corresponding possibility
of dust extinction. Because of the high redshift, only a rest-frame $B$ light curve 
could be constructed. The color derived from the difference between the
$I$ and the F850LP filters is consistent with no reddening. However, the
wavelength baseline in the rest frame is small, and the extinction to SN~1997ck
remains a major uncertainty.
The SN was discovered 11
days before maximum light in the observer's frame (Figure~1). The MLCS method, working with 
only one bandpass, finds that SN~1997ck has a normal decline rate.
Template fitting suggests that this is
a slightly over-luminous SN. 
Although the evidence that this is truly a SN~Ia is less certain for SN~1997ck
than for the other objects, the light curve looks exactly like a SN~Ia light 
curve at z = 0.97 and the pre-maximum color points in the same direction.  
But, as with some of the early observations of Perlmutter \etal\  (1997), we do not
have data to exclude conclusively other possibilities.
We treat SN~1997ck as an unreddened SN~Ia when we include it in the discussion that follows.

\begin{deluxetable}{rcccc}
\small
\tablewidth{6in}
\tablecaption{Supernova Parameters and Error Budget}
\tablehead{\colhead{} & \colhead{1995K\tablenotemark{1}} &\colhead{1997ce} &\colhead{1997cj}&\colhead{1997ck} 
}
\startdata
\sidehead{Error Budget}
CT Correction (mag) 				& 0.00 & 0.02 & 0.02 & 0.02 \nl
ZP (mag)	    				& 0.03 & 0.05 & 0.05 & 0.05 \nl
Evolution\tablenotemark{2} (mag)		& 0.06 & 0.06 & 0.06 & 0.06 \nl
Selection Bias\tablenotemark{1} (mag)		& 0.02 & 0.02 & 0.02 & 0.02 \nl
Weak Lensing\tablenotemark{3} (mag)		& 0.02 & 0.02 & 0.02 & 0.04 \nl
K-corrections\tablenotemark{4} (mag)		& 0.06 & 0.06 & 0.06 & 0.06 \nl
Light Curve Fit RPK\tablenotemark{5} (mag)	& 0.21 & 0.08 & 0.19 & 0.42 \nl  
Light Curve Fit H96\tablenotemark{5} (mag)    	& 0.13 & 0.11 & 0.11 & 0.21 \nl   
$\sigma$ of SNe~Ia (mag)			& 0.12 & 0.12 & 0.12 & 0.12 \nl	 
\sidehead{Parameters}
$z$ 					& 0.48 & 0.44 & 0.50 & 0.97 \nl
$m_B^{max}$ (mag)			& 22.89 & 22.75 & 23.19 & 24.78 \nl
$m_V^{max}$ (mag)			& 23.00 & 22.79 & 23.19 & ... \nl
$\Delta$ (mag)				&$-$0.07 & 0.41 & 0.04 & $-$0.01 \nl
$\Delta m_{15}$ (mag)			& 1.17(09) & 1.19(06) & 1.16(03) & 1.00(17) \nl
$A_V$ MLCS (mag)			&0.00 & 0.00 & 0.00 & ... \nl
Galactic ($E(B-V)$) (mag)		&0.00 & 0.01 & 0.01 & 0.01 \nl
$m-M$ RPK (mag)				&42.40(26)&41.83(18)&42.59(25)&44.15(45)\nl
$m-M$ H96\tablenotemark{6} (mag)	&42.29(20)&42.06(19)&42.48(19)&44.06(27)\nl
\enddata
\tablenotetext{1}{From SSP97} 
\tablenotetext{2}{Current observational upper limit from SSP97}
\tablenotetext{3}{From Wambsganss \etal\ 1997}
\tablenotetext{4}{Includes propagated effect on extinction, $3.1 \sigma E(B-V)$}
\tablenotetext{5}{Includes uncertainty in light curve fit, extinction and background subtraction}  
\tablenotetext{6}{Average of $B$ and $V$} 
\end{deluxetable}

\section{Discussion}

 Table~1 summarizes the error and distance estimates for this sample combined
with SN~1995K from SSP97. 
A Hubble diagram reaching to $z=1$ is shown in Figure~2. The data adhere
closely to the expectations of relativistic cosmology
as shown by the model curves. The lower panel in Figure~2 shows a detailed
comparison with the predictions of flat and open cosmological models
containing two components, nonrelativistic mater, $\Omega_m$, and a
cosmological constant, $\Omega_\Lambda$ (Caroll, Press and Turner 1994). 

\begin{figure}[ht]
\plotfiddle{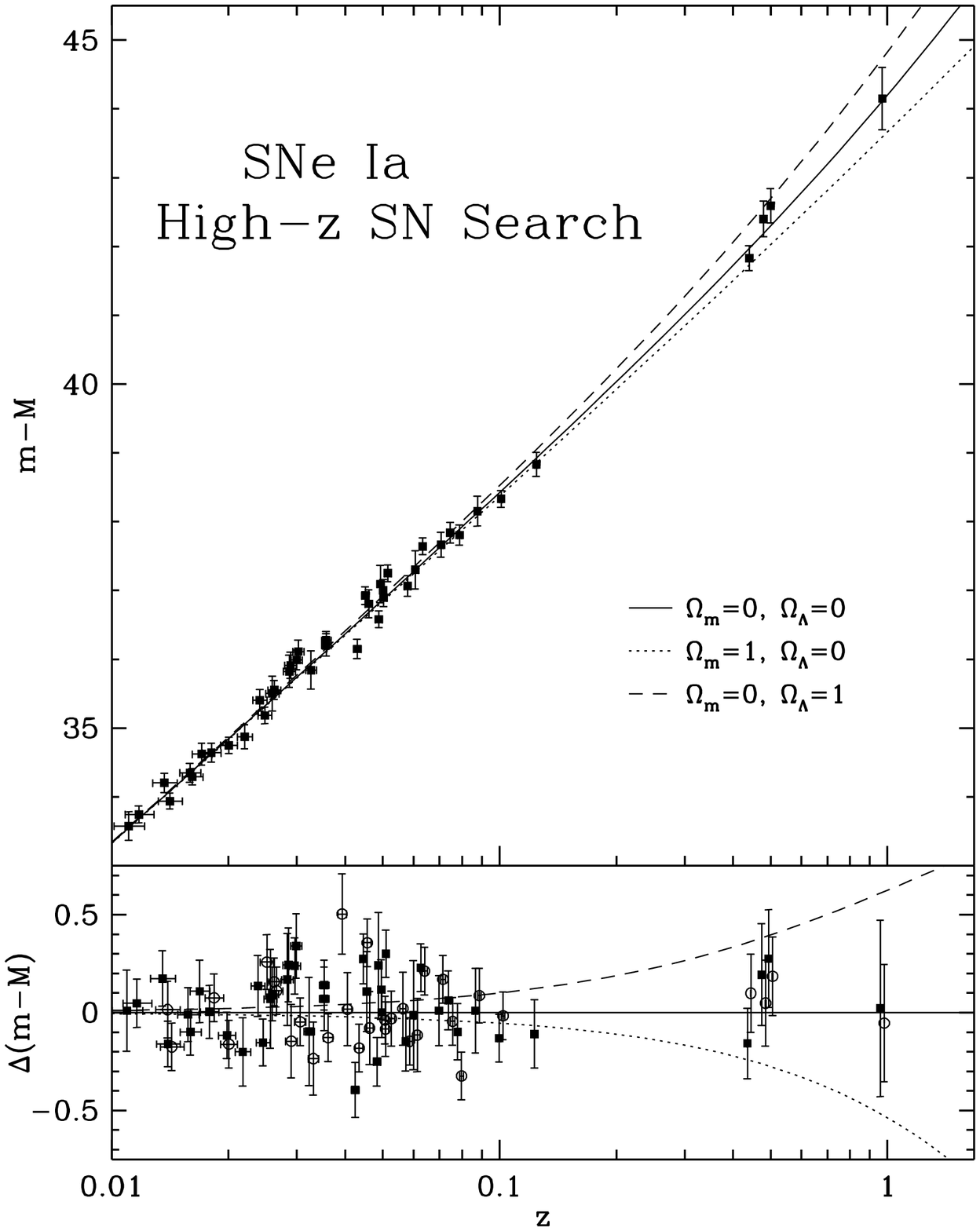}{15cm}{0.0}{80.0}{80.0}{-250.0}{-100.0}
\vspace*{2.0cm}
\caption{The Hubble diagram for SNe~Ia. The top panel shows the MLCS 
             distance modulus versus $z$ for a large sample of low redshift 
             events and the four high-$z$ SNe.
             The lower panel plots the magnitude difference between the observed 
             SNe and the magnitude expected from a Universe with
             $\Omega_m =\Omega_\Lambda = 0$. 
             The open circles show the SNe distances estimated using the 
             techniques and
             scale from H96. The redshift for SNe where both techniques have 
             been applied has been slightly shifted for clarity.}
\end{figure}

First, we will only consider the three confirmed SN~Ia which have rest frame $B$ and $V$ lightcurves.
With the constraint of a flat Universe ($\Omega_m + \Omega_\Lambda = 1$), and 
minimizing $\chi^2$ for the combined set of low-$z$ and high-$z$ SNe, we find
$\Omega_m = 0.4^{+0.3}_{-0.3}$ for the MLCS technique and 
$\Omega_m = 0.3^{+0.3}_{-0.3}$ using the template fitting method.
Alternatively, with $\Omega_\Lambda$ set to zero,
our allowed range for normal matter is  $\Omega_m = -0.1\pm 0.5$. We find that $\Omega_m < 1$
with 95\% confidence with both methods. 
Including SN~1997ck in the analysis tightens the constraints on $\Omega_M$
and $\Omega_\Lambda$, but does not significantly alter the above best fit values.
The indication from our data is that the matter density is low; as shown in Figure~3,
either the Universe is open, or if flat, then a cosmological constant makes a
considerable contribution (which may be in conflict with limits from gravitational lensing
statistics (Kochanek 1996)).
These conclusions agree with those from dynamical estimates of the
density of clustered matter (Lin \etal\ 1996, Carlberg \etal\ 1996) and from
the comparison of the Hubble time with estimates of the nuclear burning ages
of globular clusters (Reid 1997).

These results contrast with those of Perlmutter \etal\ (1997), which
preferred a high matter density even in flat models.
Although there are cosmological models acceptable to both data sets
at the 68\%\ level, the probability that the two samples have the
same parent distribution is less than 10\%.  More SNe at $z > 0.4$, colors 
and spectra for all the new objects, and a detailed comparison of the two
approaches should resolve this apparent disagreement. 

\begin{figure}[ht]
\plotone{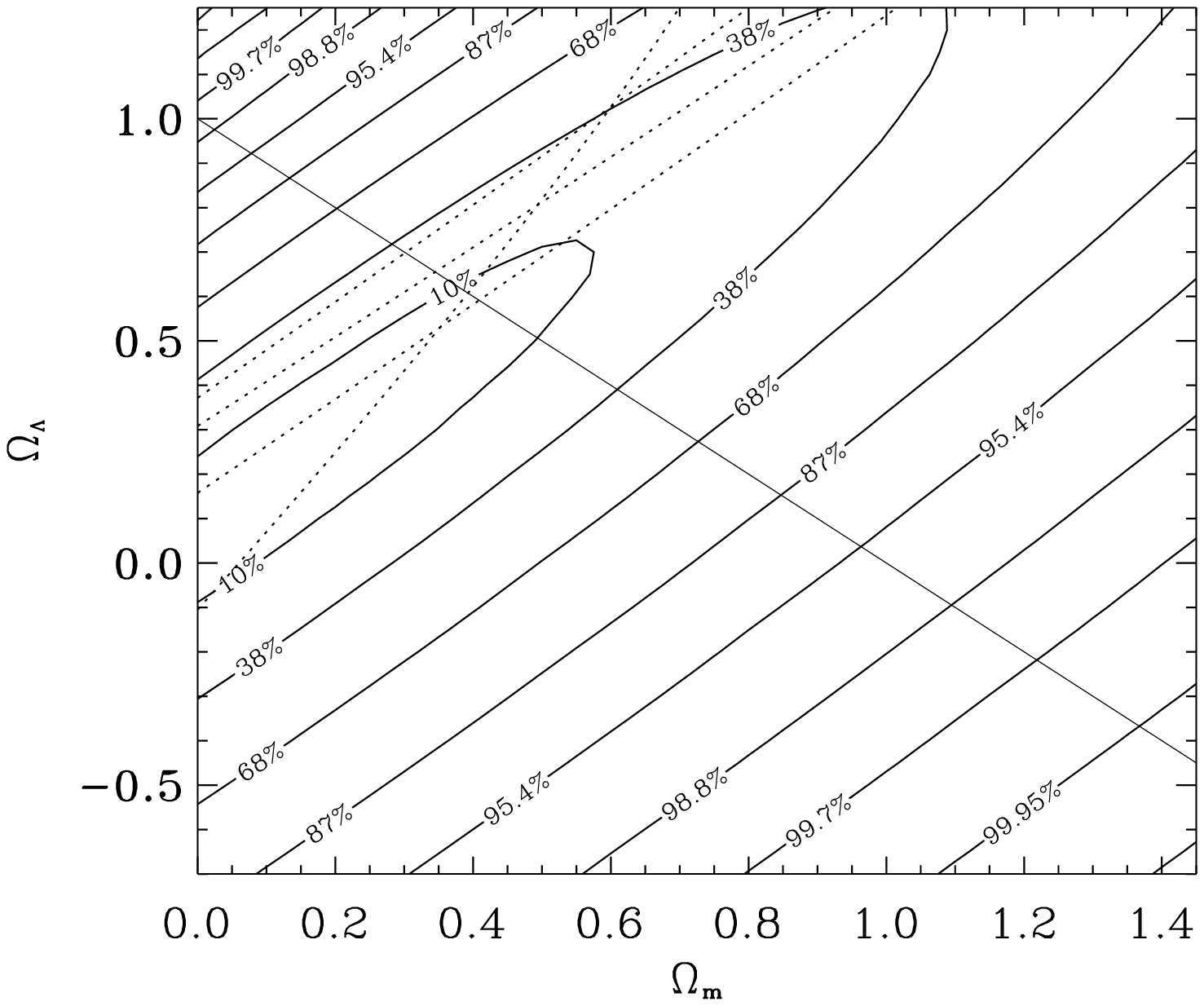}
\caption{Confidence contours in the $\Omega_m$, $\Omega_\Lambda$ plane for
             the high-$z$ SNe listed in Table~1 using the MLCS distances.
             The solid diagonal line represents the locus for a flat Universe. 
             The broken lines are the locus for individual SNe using
             distances estimated from the template fitting method.}    

\end{figure}

We have shown that SNe with redshifts as large as $z\sim 1$ can be 
discovered and successfully studied
with a combination of ground-based telescopes and \HST.
Further refinements will be made to this data set once template images are acquired with HST,
and additional ground based photometric data is obtained to determine a
more precise photometric zero point for the fields.
Our initial sample of four SNe rules out models with a high matter
density $\Omega_m \sim 1 $, although the strength  of these
conclusions should be tempered by the less-than-perfect data set for
SN~1997ck and the small size of our present sample.
Additional objects will allow us to significantly increase the precision of our measurement,
and test for sources of systematic error such as evolution.
Extending the multicolor approach to $z\sim 1$ with high precision
will be possible with NICMOS or future orbiting near-infrared instruments.

\acknowledgements

We are very grateful to STScI Director, R. Williams, for granting \HST\ director's 
discretionary time.
We thank S. Jha, G. Luppino, J. Jensen, R. Lucas, A. Patterson, D. Harmer, L. Cowie and E. Hu, 
D. Rawson, and J. Mould for their assistance.
The High-Z Supernova Team is supported by NASA through a grant from the Space Telescope ScienceInstitute,
which is operated by the Association of Universities for Research in Astronomy, Inc.,
under NASA contract NAS5-26555. CS acknowledges suppoert from the Seaver and Packard Foundations.
MH acknowledges support by the National
Science Foundation through grant number GF-1002-97 from the Association
of Universities for Research in Astronomy, Inc., under NSF Cooperative 
Agreement No. AST-8947990 and from Fundaci\'{o}n Andes under project C-12984.
Also MH acknowledges support by C\'{a}tedra Presidencial de Ciencias 1996-1997.
AC was partially supported by Fundaci\'on Antorchas Argentina under project A-13313.

\clearpage

\begin{deluxetable}{lcccc}
\tablewidth{6in}
\tablecaption{High-$z$ Supernova Photometry}
\tablehead{
\colhead{Julian Day\tablenotemark{1}} & \colhead{Telescope} &
\colhead{Filter\tablenotemark{2}} & \colhead{$B$\tablenotemark{3}}& \colhead{$V$\tablenotemark{3}}\\
}
\startdata
\sidehead{SN~1997ce}\nl
545.1 & CFHT & I            & ... & $>$25.5      \nl
568.1 & CFHT & I            & ... & 23.50 (08) \nl
573.9 & MDM & B45,V45 & 22.92 (07) & 23.09 (11)  \nl
577.0 & WIYN & R            & 22.80 (08) & ... \nl
578.0 & WIYN & I            & ... & 23.04 (07) \nl
580.5 & H88   & R,I         & 23.01 (06) & 22.83 (09)  \nl
583.3 & HST & 675W, 814W & 23.03 (02) & 22.88 (02) \nl
591.7 & HST & 675W, 814W & 23.47 (03) & 23.20 (03) \nl
598.2 & HST & 675W, 814W & 23.87 (03) & 23.57 (03) \nl
606.0 & HST & 675W, 814W & 24.58 (05) & 23.83 (03) \nl
608.7 & HST & 675W, 814W & 24.78 (05) & 23.93 (03) \nl
617.6 & HST & 675W, 814W & 25.58 (08) & 24.39 (04) \nl
\sidehead{SN~1997cj} \nl
542.8 & CFHT & I                & ... & $>$25.6      \nl
567.8 & CFHT & I                & ... & 24.03 (08) \nl
568.8 & CFHT & I                & ... & 23.73 (08) \nl
570.7 & MDM & B45,V45  & 23.44 (10) & 23.71 (13) \nl
573.8 & MDM & B45,V45  & 23.40 (08) & 23.51 (08)\nl
577.8 & KP2.1 & B45,V45   &  23.27 (14) & 23.31 (19) \nl
593.4 & HST & 675W, 814W & 23.67 (03) & 23.41 (03) \nl
594.7 & WIYN & R,I             & 23.98 (15) & 23.53 (15) \nl
596.2 & HST & 675W, 814W & 23.85 (03) & 23.58 (03) \nl
603.1 & HST & 675W, 814W & 24.36 (04) & 23.82 (04) \nl
606.1 & HST & 675W, 814W & 24.57 (05) & 23.92 (04) \nl
613.3 & HST & 675W, 814W & 25.11 (08) & 24.23 (05) \nl
\tableline
\tablebreak
\sidehead{SN~1997ck} \nl
543.1 & CFHT & I                 & $>$25.9      & ... \nl
568.0 & CFHT & I                 & 24.97 (15) & ... \nl
571.9 & MDM & V45            & 24.92 (25) & ... \nl
574.8 & MDM & Gunn-z       & 24.61 (21) & ... \nl
579.9 & H88  & I                 & 25.01 (23) & ... \nl
582.4 & HST & 850LP        & 24.77 (09) & ... \nl
591.8 & HST & 850LP        & 25.06 (10) & ... \nl
599.1 & HST & 850LP        & 25.60 (15) & ... \nl
602.1 & HST & 850LP        & 25.76 (25) & ... \nl
607.5 & HST & 850LP        & 25.63 (16) & ... \nl
620.5 & HST & 850LP        & 26.52 (19) & ... \nl
\enddata
\tablenotetext{1}{$-$2450000}
\tablenotetext{2}{B45 and V45 are defined in SSP97}
\tablenotetext{3}{Errors given in parentheses are from photon noise only} 
\end{deluxetable}

\end{document}